\documentclass[twocolumn,showpacs,aps,prl,superscriptaddress,letterpaper]{revtex4}

\usepackage{graphicx}
\usepackage{dcolumn}
\usepackage{amsmath}
\usepackage{color}
\usepackage{amssymb}

\input pubboard_babarsym_mod

\newcommand {\BzbtoDstarpennueb}  {\ensuremath{\Bzb\to\Dstarp\en\nueb}\xspace}
\newcommand {\BmtoDstarzennueb}   {\ensuremath{\Bm\to\Dstarz\en\nueb}\xspace}

\newcommand {\BzbtoDstarplmnulb}  {\ensuremath{\Bzb\to\Dstarp\ellm\nulb}\xspace}
\newcommand {\BmtoDstarzlmnulb}   {\ensuremath{\Bm\to\Dstarz\ellm\nulb}\xspace}

\newcommand {\dm}         {\ensuremath{\Delta{}m}\xspace}
\newcommand {\cosby}      {\ensuremath{\cos\theta^{*}_\mathrm{BY}}\xspace}
\newcommand {\wtilde}     {\ensuremath{\tilde{w}}\xspace}

\newcommand {\rhosqraone}     {\ensuremath{\rho^{2}_{A_1}}\xspace}

\newcommand{\FoFaF}           {\ensuremath{         F }}

\newcommand {\BrFr}[1] {\ensuremath{\mathcal{B}\left(#1\right)}\xspace}

\definecolor{Sig}{rgb}{1,1,1}
\definecolor{Dssdmp}{rgb}{0.0,1.0,1.0}
\definecolor{Dssdmf}{rgb}{0.51953125, 0.7578125, 0.63671875}
\definecolor{Cor}{rgb}{1,1,0}
\definecolor{Uncor}{rgb}{0,0,1}
\definecolor{Semisig}{rgb}{0.74609375, 0.5078125, 0.46875}
\definecolor{Dlnu}{rgb}{1.0, 0, 1.0}
\definecolor{Combds}{rgb}{0, 1, 0}
\definecolor{Ccbar}{rgb}{0.45703125, 0.5390625, 0.56640625}

\newcommand {\rd}{{\rm d}}

\newcommand {\ri}{{\rm i}}

\newcommand {\rB}{{\rm B}}

\newcommand {\rD}{{\rm D}}

\newcommand {\Begeq}{\begin{equation}}
\newcommand {\Endeq}{\end{equation}}
\newcommand {\bEa}{\begin{eqnarray}}
\newcommand {\eEa}{\end{eqnarray}}

\newcommand{\grad}{\ensuremath{^\circ}}

\newcommand{\BABARPubYear}    {07}
\newcommand{\BABARPubNumber}  {070}

\newcommand{\SLACPubNumber}   {13035}

\def\figurebox#1#2#3{%
    \def\arg{#3}%
    \ifx\arg\empty
    {\hfill\vbox{\hsize#2\hrule\hbox to #2{\vrule\hfill\vbox to #1{\hsize#2\vfill}\vrule}\hrule}\hfill}%
    \else
    {\hfill\epsfbox{#3}\hfill}%
    \fi}

\begin{document}

\preprint{\babar-PUB-\BABARPubYear/\BABARPubNumber} 
\preprint{SLAC-PUB-\SLACPubNumber} 

\begin{flushleft}
\babar-PUB-\BABARPubYear/\BABARPubNumber%
\\
SLAC-PUB-\SLACPubNumber\\
\end{flushleft}

\title{
{\large \bf
Measurement of the Decay \BmtoDstarzennueb}
}

%
%
\author{B.~Aubert}
\author{M.~Bona}
\author{Y.~Karyotakis}
\author{J.~P.~Lees}
\author{V.~Poireau}
\author{X.~Prudent}
\author{V.~Tisserand}
\author{A.~Zghiche}
\affiliation{Laboratoire de Physique des Particules, IN2P3/CNRS et Universit\'e de Savoie, F-74941 Annecy-Le-Vieux, France }
\author{J.~Garra~Tico}
\author{E.~Grauges}
\affiliation{Universitat de Barcelona, Facultat de Fisica, Departament ECM, E-08028 Barcelona, Spain }
% ...
\collaboration{The \babar\ Collaboration}
\noaffiliation

\date{\today}%

\vfill\newpage %

\begin{abstract}
Using 226 million \BB events recorded on the
\FourS resonance with the \babar{} detector at the SLAC \epem
\pep2 storage rings,
we reconstruct \BmtoDstarzennueb decays using the decay chain
$\Dstarz\to\Dz\piz$ and $\Dz\to{}\Km\pip$.
From the dependence of their differential rate on $w$, the dot
product of the four-velocities of  \Bm and \Dstarz{},  
and using the form factor description by Caprini {\it{et al.}}
with the parameters $\FoFaF(1)$ and \rhosqraone,
we obtain the results 
$\rhosqraone        = 1.16\pm 0.06 \pm 0.08$, 
$\FoFaF(1)\cdot\Vcb =(35.9\pm 0.6  \pm 1.4)\cdot 10^{-3}$, and 
$\BrFr{\BmtoDstarzennueb} = (5.56\pm 0.08\pm 0.41)\%$.

\end{abstract}


\maketitle

The Standard Model of particle physics (SM) contains a large number of free
parameters which can only be determined by experiment. Precision
measurements of all of these parameters are essential for probing the
validity range of the model by comparing 
many other
precision measurements with SM calculations. 
One of the SM parameters, the element \Vcb of the Cabibbo-Kobayashi-Maskawa
quark-mixing matrix, is determined with semileptonic $B$-meson decays.
Their rates $\Gamma$ are given by the universality of the weak
interaction (the Fermi constant $ G_{\rm F}$), by quark mixing 
($\Gamma \propto G_{\rm F}^2 {\Vcb }^2$), 
and by strong-interaction corrections calculated in heavy-quark
effective QCD. 
For the exclusive decays 
\BzbtoDstarplmnulb{} and \BmtoDstarzlmnulb{}
($\ell=\electron,\mmu$),
these corrections are expressed as form factors in the differential
rate ${\rm d}\Gamma/ {\rm d}w$, where $w$ is the dot product of the
four velocities of the $B$ and the $\Dstar${}. 
The form factors depend on the three
parameters $\rho^2$, $R_1(1)$, and
$R_2(1)$ \cite{Caprini:1997mu}. 
Whereas the \Bzb mode has been measured by many
experiments \cite{Barberio:2007cr}, the \Bm mode has only been
measured by two groups \cite{Adam:2002uw,Albrecht:1991iz} with much
smaller data samples.
However, the \Bzb experiments do not agree well in their $\rho^2$
results. Using the isospin symmetry 
${\rm d}\Gamma (\BmtoDstarzlmnulb) = {\rm d}\Gamma (\BzbtoDstarplmnulb)$,
a precision measurement of the \Bm mode can improve knowledge of $\rho^2$
and consequently of $\Gamma$ and \Vcb.

The aim of our analysis \cite{JensThesis} is the determination of
the differential decay fraction $\rd{\cal B}(\BmtoDstarzennueb)/\rd w$, 
where ${\cal B} = \Gamma \tau$, with the \Bm lifetime $\tau${}.
The neutrino in the \BmtoDstarzennueb decay is not reconstructed.
Therefore, the $w$ value of each reconstructed event cannot be obtained, 
only an approximation
\wtilde as defined below. 
Instead of unfolding $\rd{\cal B}/\rd {\wtilde}${}, the parametrized
$\rd{\cal B}/\rd w$ expectation convolved with the $w$ resolution from
Monte Carlo (MC) simulation is fitted to the observed $\rd{\cal B}/\rd {\wtilde}$
distribution. 
The fit uses the parametrization of Caprini et al.\
\cite{Caprini:1997mu}
with $\rho^2\equiv\rho^2_{A_1}$
and determines the two parameters $\FoFaF(1)\cdot\Vcb$ and $\rho^2$.
The decay fraction ${\cal B}$
is obtained by integrating $\rd{\cal B}/ \rd w$. 
Using the notations 
$\Delta M \equiv m_B-m_{D^*}$, 
$r \equiv m_{D^*}/m_B$, and
$z \equiv (\sqrt{w+1}-\sqrt{2})/(\sqrt{w+1}+\sqrt{2})$, 
the parametrization is defined by the following expressions: 
\begin{gather*} 
\frac{\rm d\Gamma}{{\rm d}w}= 
   \frac{ G_{\rm F}^2 {\Vcb }^2} {48{\pi}^3}
   (\Delta M)^2 m_{\Dstar}^3 \sqrt{w^2-1} \left( w+1 \right)^2 g(w) \FoFaF^2(w) ,\\
g(w)= 1 + \frac{4w}{w+1} \frac{m_{B}^2-2w m_{B} m_{\Dstar}+ m_{\Dstar}^2}{(\Delta M)^2} ,\\
F^2(w)= \frac{\left| h_{A_1}(w) \right|^2}{g(w)}
   \sum_{i=0,+,-}  \left|  \tilde{H}_i(w)\right|^2 ,\\
\tilde{H}_0(w)=1+ \frac{w-1}{1-r} \left[1-R_2(w)\right] ,\\
\tilde{H}_{\pm} (w)=\frac{\sqrt{1-2wr+r^2}}{1-r}
   \left[ 1 \mp \sqrt{\frac{w-1}{w+1}}R_1(w) \right]
\end{gather*} 
\begin{gather*} 
\frac{h_{A_1}(w)}{h_{A_1}(1)} = 1-8\rho^2 z
   +\left(53\rho^2-15\right)z^2-\left(231\rho^2-91\right)z^3 ,\\
R_1(w)  =  R_1(1) - 0.12(w-1) + 0.05(w-1)^2,\\
R_2(w)  =  R_2(1) + 0.11(w-1) - 0.06(w-1)^2,
\end{gather*}
with $F(1)=h_{A_1}(1)$. 
The values of $R_{1,2}(1)$ are not determined in this analysis; they
are taken from Ref.~\cite{Aubert:2007rs}.

For our analysis, we use 205\invfb of \epem annihilation data recorded at 
$\sqrt{s}\approx{}m(\FourS)$ with the \babar{} detector
\cite{Aubert:2001tu}
at the SLAC PEP-II storage rings \cite{PepII}.
In addition to these on-peak data, we also use 16\invfb of off-peak data
collected 40\mev below the \FourS resonance.
We select \BmtoDstarzennueb candidates \cite{Charge_Conj_Decays}
by pairing electrons with $p^*>1.2${}\gevc in the \epem rest frame (cms)
with \Dstarz candidates.
Since the precision of our results is
not statistically limited, we restrict the analysis
to the sequential decay modes $\Dz\to\Km\pip${}, which has the smallest
combinatorial background, and $\Dstarz\to\Dz\piz${}, which
has a better resolution in
$\dm\equiv{}m(\Km\pip\piz)-m(\Km\pip)$ than
$\Dstarz\to\Dz\gamma${}.

Charged particles are selected if they have at least 10 hits
in the drift chamber, transverse momentum $\pt>0.1${}\gevc, and a polar 
angle between
23.5\grad{} and 145.5\grad{} in the laboratory frame. 
Electrons (kaons) are selected with tight 
(loose) particle identification criteria 
\cite{Kaons_Electrons_BABAR}. Neutral pions are reconstructed from two
photons, each with energy above 30\mev and a photon-compatible lateral
shower shape in the calorimeter. 
The two photons must be consistent with
the \piz hypothesis ($115 < m_{\gamma\gamma} < 150\mevcc$). A
kinematic fit with the constraint $m_{\gamma\gamma}=m_{\pi^0}${}
improves the \dm resolution by a factor of 3.
The decay candidates have to fulfill
the following additional requirements:
the \Dstarz{}-\Dz mass difference and the \Dz-candidate mass
must satisfy $135<\Delta m<153\mevcc$ and
$1.8496<m(\Km\pip)<1.8796\gevcc$, respectively.
To reject non-$B${}-decay candidates, the second normalized Fox-Wolfram moment
\cite{Fox:1978vw} of the event
has to be smaller than 0.45. To help reject combinatorial background
with a \Dstarz and an \en from different \B mesons in the event, the
cms angle between them must be larger than 90\grad.

Since there are many low-energy background photons, the selection criteria result
in many events with two or more
$\Dstarz\electron$ candidates, on average 1.75 per event. 
All $\Dstarz\electron$ candidates
in the same $\electron{}K\pi$ combination form one group, called
a candidate group. On average there are 1.015 candidate groups per
event. 
When an event has more than
one candidate group, we keep only the one with the
best $|m(K\pi)-m(\Dz)|$. All
candidates in one group are kept in the analysis because the simulation of
low-energy photons is not perfect.
This procedure ensures that correctly reconstructed candidates are
selected with the same probability in data and MC simulation.

The surviving candidates are binned in \dm, \cosby, and \wtilde. 
The first two variables are used for
signal-background separation, and the third is used for
the $w$ dependence of the signal. The mass difference \dm is defined above, and
$\theta_{\rm BY}^*$ is the angle between 
the \B meson and the $Y=\Dstarz+e$ system in the cms defined by
$$p_\nu^2=0=m_\rB^2+m_{\rm Y}^2-2(E_\rB^* E_{\rm Y}^*
                  -|{\vec p}_\rB^{\,*}||{\vec p}_{\rm Y}^{\,*}|\cos\theta_{\rm BY}^*)\ .$$
The value of
$$
w=w(\beta^*)\equiv
\left({E_\rB^* E_{\rD^*}^*-|{\vec p}_\rB^{\,*}||{\vec p}_{\rD^*}^{\,*}|\cos\beta^*}\right)
/ 
\left({m_\rB m_{\rD^*}}\right)
$$
cannot be determined since the angle $\beta^*$ between the \B and the \Dstarz
in the cms is
unknown. However, $\beta^*$ is bounded by a minimum and a maximum value and we use
${\tilde w}=[w(\beta^*_{\rm min})+w(\beta^*_{\rm max})]/2$
as an estimator for $w$. 
Both $w$ and $\tilde w$ range from 1.0 to
1.5, and the distribution of ${\wtilde}-w$ is
nearly Gaussian with an RMS of  0.026.

The fit for 
$V=\FoFaF(1)\Vcb$ and $\rho^2$
is a binned maximum-likelihood fit
with 41, 14, and 10 equidistant bins in \dm, \cosby, and \wtilde, respectively.
The fit function in each $\wtilde${} bin is the sum
of the signal function $S_{\wtilde}(V,\rho^2)$ and 23
background functions $B_{i,\wtilde}(V,\rho^2)${}. Each summand is 
taken as the product of one-dimensional functions of \dm and \cosby.
The \dm distributions of correctly (wrongly) reconstructed \Dstarz mesons are
parametrized by the sum of 3 bifurcated Gaussians (product of an
exponential and a power law function). The \cosby distributions are
modeled by modified KEYS functions \cite{JensThesis}.

The factor functions of $S_{\wtilde}$
are obtained from fits to the reweighted
signal MC distributions with
$V$-, $\rho^2$-, $R_1(1)$-, and $R_2(1)$-dependent weights on the generator level.
$S_{\wtilde}$ also
includes the total number of produced \BB pairs,
all decay fractions of sequential decays, the \Bm lifetime, all MC
reconstruction efficiencies, and efficiency corrections. 
The corrections for track reconstruction and charged-particle
identification are obtained
from control data samples and their MC expectations.
The correction of the \piz reconstruction efficiency is described below.
Small corrections are also applied for deviations of the shapes of the \dm distributions
in data and MC
because of track resolution differences, and for deviations in the shapes of
the \cosby distributions because of differences in storage-ring energy calibration and resolution.

The background functions are separately determined
for the 23 background classes \cite{JensThesis}.
The large number of backgrounds is necessary in order
to factorize all $B_{\ri,{\tilde w}}$ 
as $B_{1,\ri,\wtilde}(\dm)\times B_{2,\ri,\wtilde}(\cosby)$.
The one-dimensional fit functions $B_{j,\ri,\wtilde}$ are again
obtained from fits to MC distributions.
The fit to the data has 49 free parameters; $V${}, $\rho^2${}, and 47 for adjustments of \dm shapes, \cosby shapes, and background fractions. 
The number of $\epem\to\ccbar$ background events is fixed by the off-peak
data.

As validation of the fit procedure, we perform our fit on five different MC
subsamples whose size corresponds to that of the data sample.
All five results for $V$ and $\rho^2$
agree with the MC input to within one standard deviation. 
Applied to the data and using the input-parameter values in
Table~\ref{tbl:result:input_parameters}, 
\begin{table}
\begin{center}
\caption
{Summary of input parameter values. }
\begin{tabular}{lll}
\hline\hline
{{Input Parameter}}&
{{Value}}&
{{Ref.}}\\
\hline
$\BrFr{\FourS\to\BpBm}$      &  $(50.6 \pm 0.8)\%$     & \cite{pdg:2006} \\
$\BrFr{\Dstarz\to\Dz\piz}$   &  $(61.9 \pm 2.9)\%$     & \cite{pdg:2006} \\
$\BrFr{\Dz\to\Km\pip}$       &  $(3.80 \pm 0.07)\% $   & \cite{pdg:2006} \\
$\BrFr{\piz \to \gaga}$      &  $(98.798 \pm 0.032)\%$ & \cite{pdg:2006} \\
$\tau_{\Bm}$                 &  $(1.638\pm 0.011)\ps$  & \cite{pdg:2006} \\
$R_1(1)$                     &  $1.429 \pm 0.075$      & \cite{Aubert:2007rs}\\
$R_2(1)$                     &  $0.827 \pm 0.044$      & \cite{Aubert:2007rs}\\
\hline\hline
\end{tabular}
\label{tbl:result:input_parameters}
\end{center}
\end{table}

the fit result is $V=(35.9\pm 0.6)\cdot 10^{-3}$ and $\rho^2=1.16\pm 0.06$ with
a correlation coefficient of +0.90. 
The result leads to  ${\cal B}=(5.56\pm 0.08)\%$ after integrating
$\rd{\cal B}/\rd w${}. 
The total number of signal events is $23\,499\pm329$.
A control value of $\chi^2$ can be calculated after the fit as
a goodness-of-fit measure. We find 4436.3 for 4095 degrees of freedom
after rebinning in regions with low statistics.
The values of $\chi^2$ in the MC-subsample fits
are of similar size indicating that the factorization assumptions
for $S_{\tilde w}$ and $B_{i, \tilde w}$ are not perfect.
Since there is no bias in $V$ or $\rho^2$ in the MC-subsample fits and no significant correlation
between background parameters and both $V$ and $\rho^2$ in the fit to the data, we conclude 
that the results are unbiased.

\begin{figure}
\begin{center}
\includegraphics[width=0.40\textwidth]{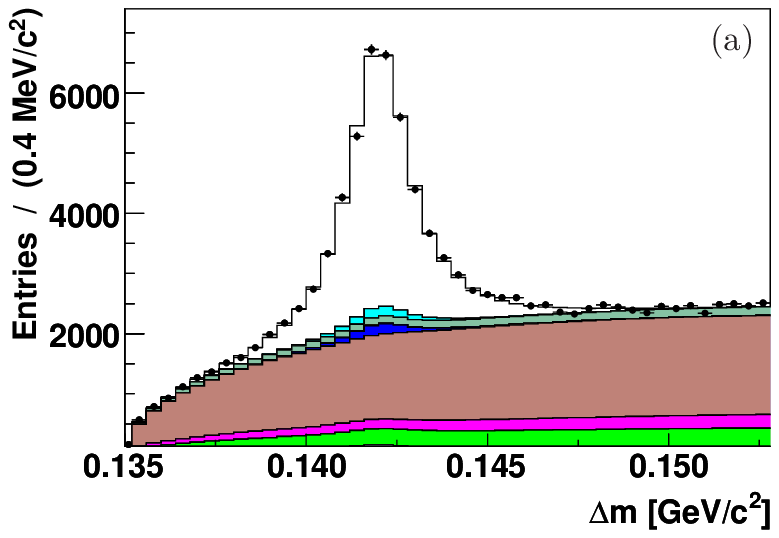}
\includegraphics[width=0.40\textwidth]{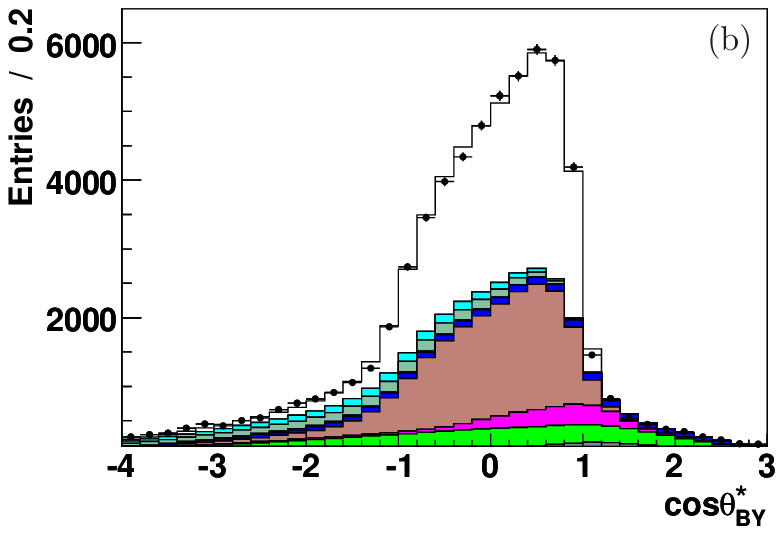}
\includegraphics[width=0.42\textwidth,clip]{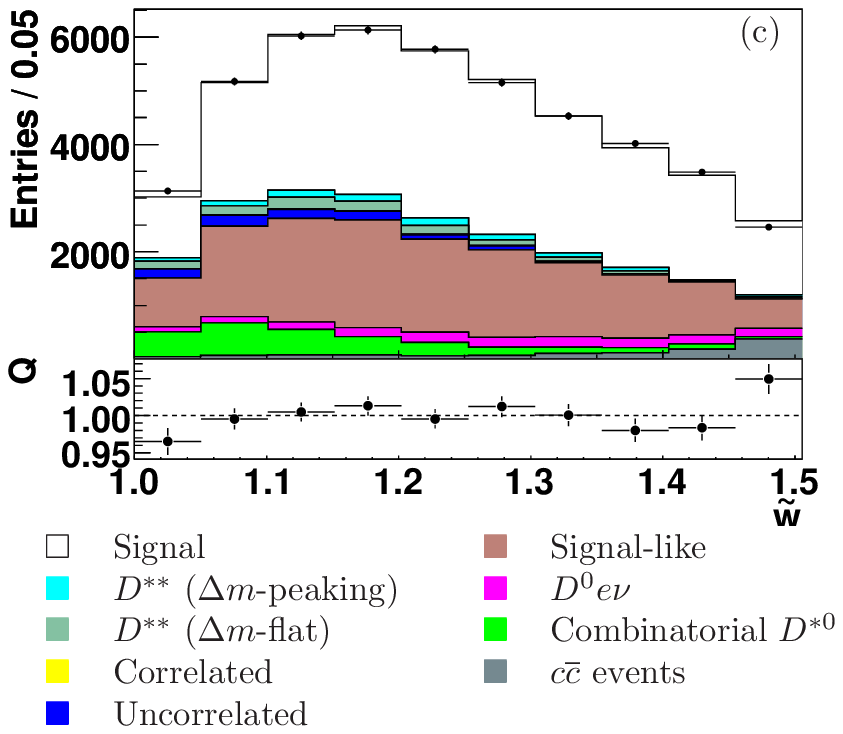}
\caption
{
(Color online). 
Data distributions (dots with error bars) and fit results
(stacked histograms) for (a) \dm in the \cosby signal range ($-$1,+1), 
(b) \cosby in the \dm signal range (140,144\mevcc), and
(c) \wtilde in both signal ranges. The plot below (c) shows the quotient fit/data.
The different contributions to the fit function are explained in the
text.
}
\label{fig:dm_cosby_wtilde}
\end{center}
\end{figure}
Figure~\ref{fig:dm_cosby_wtilde} shows the result of the fit together with the selected data. The ``Signal'' 
part of the fit function contains the correctly reconstructed 
\BmtoDstarzennueb decays. The two $D^{**}$ parts contain 
$B\to{}D^{**}\electron\nu$ decays with (``\dm peaking'') and without (``\dm flat'') a
correctly reconstructed \Dstarz intermediate state 
($D^{**}=D_{1},D^{*}_{0},{D'}_{1},D^{*}_{2},D^{*}\pi,D\pi$). 
Events with a correctly reconstructed
\Dstarz and a correctly identified electron from the same \B and from
two different \B mesons are in the ``Correlated'' and ``Uncorrelated''
background parts, respectively. ``Signal-like'' are true decays
\BmtoDstarzennueb and \BzbtoDstarpennueb which are not correctly
reconstructed. The background from true $\B\to\Dz\electron\nu$ decays
is called ``$\Dz\electron\nu$''.  All other background candidates from
\BB events (``Combinatorial \Dstarz{}'') are flat in the \dm and the
\cosby distributions since they do not contain a correctly
reconstructed \Dstarz and they do not come from a charmed semileptonic
decay. 
The last contribution, only visible at high 
\wtilde, comes from \ccbar events.

To determine the systematic uncertainties listed in
Table~\ref{tbl:systematics:all} we either rerun the fit 
with varied input or we rescale the fit result. The upper part of the Table
gives the ``internal'' uncertainties which are specific to our analysis.
The relative uncertainty on the efficiency to reconstruct a track is
0.8\%, leading to 2.4\% and 1.2\% for ${\cal B}$ and $V$.
The dependence of the tracking efficiency on the transverse momentum
\pt has an uncertainty which could distort the shape of the \wtilde
spectrum.%
\begin{table}
\begin{center}
\caption
{Relative systematic uncertainties in percent.}
\begin{tabular}{lccc}
\hline
\hline
\\[-2.4ex]
\ &
${{\Delta{}V}/{V}}$ &
${{\Delta\rho^2}/{\rho^2}}$ &
${{\Delta\cal B}/{\cal B}}$ \\
\\[-2.4ex]
\hline
Tracking efficiency ($\epsilon_{\rm tr}$)&
1.2 &
-   &
2.4 \\
\pt dependence of $\epsilon_{\rm tr}$&
0.3 &
0.5 &
0.2 \\
Particle ID efficiency&
0.9 &
2.0 &
1.6 \\
Extrapolated \piz efficiency ($\epsilon_{\piz}$)&
1.8 &
- &
3.6 \\
$p_{\piz}$ dependence of $\epsilon_{\piz}$&
1.0 &
3.5 &
0.4 \\
\dm shape of $D^{**}${} background &
0.1 &
0.1 &
0.2 \\
Shape parameters&
1.0 &
2.5 &
0.6 \\
Number of \BB events&
0.6 &
- &
1.1 \\
Off-peak luminosity &
0.1 &
0.4 &
$<$0.1 \\
MC statistics&
0.3 &
0.8 &
0.2 \\
Radiative corrections&
0.5 &
0.4 &
1.4 \\
\hline
{Total internal} &
${2.9}$ &
${4.9}$ &
${5.0}$ \\
\hline
$R_1(1)${} and $R_2(1)$ &
0.4 &
4.7 &
0.5 \\
$\BrFr{\FourS\to\BpBm}${} &
0.8 &
- &
1.6 \\
$\BrFr{D^{*0} \to \Dz \piz}${} &
2.3 &
- &
4.7 \\
$\BrFr{\Dz \to K^- \pi^+}${} &
0.9  &
- &
1.8 \\
$B^{-}${} life time &
0.3 &
- &
- \\
$D^{**}${} decay fractions &
0.3 &
0.7 &
0.3 \\
Number of \Dstarz in \ccbar events&
0.2 &
0.7 &
$<$0.1  \\
\hline
{Total external} &
${2.7}$ &
${4.8}$ &
${5.3}$ \\
\hline 
{Total } &
${3.9}$ &
${6.8}$ &
${7.3}$ \\
\hline 
\hline 
\end{tabular}
\label{tbl:systematics:all}
\end{center}
\end{table}

The uncertainties arising from the identification (ID) of charged tracks
as electrons or as kaons contribute to the result as listed under
``particle ID efficiency''.
A significant fraction of the total uncertainty comes
from the precision of the \piz reconstruction efficiency
($\epsilon_{\piz}$). It is
determined from $\epem\to\tautau$ events where one of the two $\tau$
leptons is either reconstructed
by one track and two clusters (mainly $\tau\to\rho(\pi\piz)\nu$) or 
by only one track without clusters (mainly
$\tau\to\pi\nu,\mu\nu\nub$). 
The other $\tau$, used as a $\tau${}-pair tag, 
is reconstructed in its $\electron\nu\nub$ decay.
From the numbers of \tautau events reconstructed in each of the two
channels we derive an efficiency in data and in MC,
giving a correction to the simulated \piz efficiency.
The correction is obtained for momenta above 350\mevc and has a
precision of 3\%. 
In the lower-momentum region with all \piz mesons from $\Dstarz\electron\nu$ decays,
we use a correction factor of  $0.960\pm 0.035$ where the increased uncertainty covers the
extrapolation into this region.
Efficiency differences between \tautau and \BB events are covered by 
the MC simulation as controlled by comparing the rates
of reconstructed \Dz decays into \Km\pip and  \Km\pip\piz.
The uncertainty
in the shape of the \wtilde spectrum, i.e. its influence on $\rho^2${}, is
estimated by fit results for different lower cuts on $p_{\piz}$
(``$p_{\piz}$ dependence of $\epsilon_{\piz}$'').
Corrections to the \dm shape and to the \cosby shape are parametrized
as functions of \wtilde, see ``shape parameters'' for their
contributions to the systematics.
Uncertainty estimates from radiative corrections are taken from the \babar\ analysis 
of $B^0\to D^*\ell\nu$ decays
\cite{Aubert:2007rs}
which uses the same lepton-momentum cutoff of 1.2 GeV/c.

The ``external'' uncertainties owing to parameters taken from other experiments are given
in the lower part of  Table~\ref{tbl:systematics:all}. For $\rho^2$ 
they are dominated by $R_1(1)$ and $R_2(1)$. 
For future updates, we also give in Table~\ref{tbl:deriv} 
the derivatives of  our three results with respect to these two variables as determined from 
fits with varied input values.
\begin{table}
\begin{center}
\caption
{Derivatives of $V$, $\rho^2$, and ${\cal B}$.}
\begin{tabular}{lccc}
\hline\hline
\\[-2.4ex]
\phantom{VVVVVVV} & 
$V$  &
$\rho^2  $        &
${\cal B}$       \\
\hline
\\[-2.0ex]
$\partial{}/\partial{}R_1^{\phantom{I}}(1)$ & 
 $-0.00342$     &
 $+0.0303$  &
 $-0.00567$\\
$\partial{}/\partial{}R_2(1)$ &
$-0.00525$   &
$-1.22$ &
$-0.00594$ \\
\hline\hline
\end{tabular}
\label{tbl:deriv}
\end{center}
\end{table}
The $\B \to D^{**} e \nu$ decays contribute to the
uncertainties because of their less precisely known decay fractions
and their 
uncertain \dm shape due to low-energy photon background.
Uncertainties in their \wtilde shape are covered 
by 10 of the 49 fit parameters.

Adding all systematic uncertainties in quadrature leads to the last line in
Table \ref{tbl:systematics:all} and to our final results
\begin{gather*}
\phantom{wwww}\FoFaF(1)\cdot\Vcb=(35.9\pm 0.6\pm 1.4)\cdot 10^{-3}\ ,\\
\phantom{wwww}\phantom{ii}\rhosqraone= 1.16\pm 0.06\pm 0.08\ , \\
{\cal B}(B^-\to D^{*0} e^-\overline\nu_e) = (5.56\pm 0.08\pm 0.41)\%\ .\phantom{ww}
\end{gather*}
The correlation coefficients between $\FoFaF(1)\cdot\Vcb$ and 
$\rhosqraone$ are +0.90 for statistics, +0.42 for systematics,
and +0.52 in total. 
Using $F(1)=0.919\pm 0.033$ from lattice QCD
\cite{Hashimoto:2001nb},
we obtain $\Vcb=(39.0\pm 0.6\pm 2.0)\cdot 10^{-3}$ in good agreement
with the average from the exclusive neutral \B decays $\Bz\to\Dstarm\ellp\nu$,
$(39.2\pm 0.7\pm 1.4)\cdot 10^{-3}$
\cite{Barberio:2007cr},
and in agreement with results from the inclusive decays $\B\to{}X_{c}\ell\nu$,
e.\ g.\ $(42.0\pm 0.2\pm 0.7)\cdot 10^{-3}$ in
Ref.\ \cite{Buchmueller}. 
Our result for
$\rho^2$ is in the center of the range (0.5, 1.5) from the $\Bz\to\Dstarm\ellp\nu$
experiments \cite{Barberio:2007cr}.

Compared with the PDG average \cite{pdg:2006} of 
${\cal B}(B^-\to D^{*0} e^-\overline\nu_e)$, our result is lower by more than 
1.5 standard deviations.
For a comparison of our decay-fraction result with that of the
$B^0$ mode, we use 
$\tau(\Bp)/\tau(\Bz)=1.076\pm 0.008$ and  
${\cal
B}(\Bz\to\Dstarm\ellp\nu)=(5.28\pm{}0.18)\%${} \cite{Barberio:2007cr}. 
This gives 
${\cal B}(\Bm\to\Dstarz\ellm\overline\nu)=(5.68\pm 0.20)\%$;
our result agrees well with this value.

To conclude, this measurement is the first one for \BmtoDstarzlmnulb
decays with a data
sample comparable to recent \BzbtoDstarplmnulb experiments. The results
for the decay
rate and for \Vcb agree well with the \Bzb mean values. Since the
uncertainties in the
reconstruction of low-momentum \pip and \piz are experimentally very
different, the
agreement of our $\rho^2${} result with the central value of the \Bzb results
provides a crucial cross check for previous
\Vcb determinations in $\B \to \Dstar \ell \nul$ decays.

We are grateful for the excellent luminosity and machine conditions
provided by our \pep2\ colleagues, 
and for the substantial dedicated effort from
the computing organizations that support \babar.
The collaborating institutions wish to thank 
SLAC for its support and kind hospitality. 
This work is supported by
DOE
and NSF (USA),
NSERC (Canada),
CEA and
CNRS-IN2P3
(France),
BMBF and DFG
(Germany),
INFN (Italy),
FOM (The Netherlands),
NFR (Norway),
MIST (Russia),
MEC (Spain), and
STFC (United Kingdom). 
Individuals have received support from the
Marie Curie EIF (European Union) and
the A.~P.~Sloan Foundation.



\begin{thebibliography}{99}

\bibitem{Caprini:1997mu}
  I.~Caprini, L.~Lellouch and M.~Neubert,
  Nucl.\ Phys.\ B {\bf 530}, 153 (1998).

\bibitem{Barberio:2007cr}
  E.~Barberio {\it et al.}  [HFAG Collaboration],\\
  arXiv:0704.3575 [hep-ex].

\bibitem{Adam:2002uw}
  N.~E.~Adam {\it et al.}  [CLEO collaboration],
  Phys.\ Rev.\ D {\bf 67}, 032001 (2003).

\bibitem{Albrecht:1991iz}
  H.~Albrecht {\it et al.}  [ARGUS Collaboration],
  Phys.\ Lett.\ B {\bf 275}, 195 (1992).

\bibitem{JensThesis}
  For further details see J.\ Schubert, TU Dresden Dr.\ rer.\ nat.\ Thesis 2006, 
  SLAC-R-876 (2007).


\bibitem{Aubert:2007rs}
  B.~Aubert {\it et al.}  [BABAR Collaboration],\\
  arXiv:0705.4008 [hep-ex], submitted to Phys.\ Rev.\ D.

\bibitem{Aubert:2001tu}
  B.~Aubert {\it et al.}  [\babar{} Collaboration],
  Nucl.\ Instrum.\ Meth.\  A {\bf 479}, 1 (2002).


\bibitem{PepII} 
   PEP-II Conceptual Design Report,
   SLAC-418 
   (1993).

\bibitem{Charge_Conj_Decays}
  The use of charge-conjugated states in this analysis is always implied. 

\bibitem{Kaons_Electrons_BABAR}
  B.~Aubert {\it et al.}  [BABAR Collaboration],
  Phys.\ Rev.\  D {\bf 67}, 031101 (2003)
  and
  J.~Schwiening {\it et al.}  [BABAR-DIRC Collaboration],
  Nucl.\ Instrum.\ Meth.\  A {\bf 553}, 317 (2005).

\bibitem{Fox:1978vw}
  G.~C.~Fox and S.~Wolfram,
  Nucl.\ Phys.\ B {\bf 149}, 413 (1979)
  [Erratum-ibid.\ B {\bf 157}, 543 (1979)].

\bibitem{pdg:2006}
  W.-M.~Yao {\it et al.}  [Particle Data Group],
  J.\ Phys.\ G\ {\bf 33}, 1 (2006).


\bibitem{Hashimoto:2001nb}
  S.~Hashimoto {\it et al.}, 
  Phys.\ Rev.\ D {\bf 66}, 014503 (2002).


\bibitem{Buchmueller} O. Buchm\"uller and H. Fl\"acher, Phys. Rev. D {\bf 73}, 073008 (2006).

\end{thebibliography}
\end{document}